\documentclass[letterpaper,twocolumn,10pt]{article}
\usepackage{usenix}
\usepackage{enumitem} 
\usepackage{graphicx}
\usepackage{hyperref}
\hypersetup{colorlinks=true}
\usepackage{pifont}
\usepackage{subcaption}
\usepackage{xcolor}
\usepackage{booktabs}
\usepackage{multirow}
\usepackage{amsmath} 

\usepackage{amssymb}
\usepackage{amsfonts}
\usepackage{mathtools}
\usepackage{amsthm}
\usepackage{mathrsfs}
\usepackage{newtxmath} 
\usepackage{xspace}
\usepackage{cleveref}
\usepackage{xcolor}
\usepackage{minted}
\usepackage{listings}
\usepackage{comment}
\usepackage{inconsolata}

\usepackage{titlesec}               % change \paragraph
\titleformat{\paragraph}[runin]     % same line as text
  {\normalfont\normalsize\bfseries} % Use normal (non-italic), normal size, bold font
  {}                                % No numbering
  {0pt}                             % No space between label (empty) and title
  {}[]                             % Period after title
\titlespacing*{\paragraph}
  {0pt}                             % No left indentation for title
  {0.5\baselineskip}                % Vertical space before the title
  {1em}                             % Hspace after title before text starts

\makeatletter
\def\@maketitle{\newpage
 \vbox{ % to 2.5in{
 \begin{center}%
  {\Large\bf \@title \par}%
  \vskip 0.3cm
  {\large\it
   \lineskip .5em
   \begin{tabular}[t]{c}\@author
   \end{tabular}\par}%
 \end{center}%
 \par
 }
}

\setlist[itemize]{noitemsep, topsep=0pt, parsep=0pt, partopsep=0pt}

\newtheorem{theorem}{Theorem}

\newcommand{\system}{TTrace\xspace}
\newcommand{\bug}{silent bug\xspace}
\newcommand{\bugs}{silent bugs\xspace}
\newcommand{\mch}{\text{mch}}

\newcommand{\code}[1]{\texttt{\detokenize{#1}}}

\newcommand{\yes}{\ding{51}}
\newcommand{\no}{\ding{55}}

\newcommand{\Order}{\mathcal{O}}
\newcommand{\E}{\mathbb{E}}

\newcommand{\norm}[1]{\left\lVert#1\right\rVert}

\begin{document}

\title{\system: Lightweight Error Checking and Diagnosis for Distributed Training}

\author{
Haitian Jiang \\
\small New York University \\
\small\texttt{haitian.jiang@nyu.edu}
\and
Shaowei Zhu \\
\small Amazon Web Services \\
\small\texttt{shaowz@amazon.com}
\and
Zhen Zhang \\
\small Amazon Web Services \\
\small\texttt{zhzhn@amazon.com}
\and
Zhenyu Song \\
\small Amazon Web Services \\
\small\texttt{zhenyus@amazon.com}
\and
Xinwei Fu \\
\small Amazon Web Services \\
\small\texttt{fuxinwe@amazon.com}
\and
Zhen Jia \\
\small Amazon Web Services \\
\small\texttt{zhej@amazon.com}
\and
Yida Wang \\
\small Amazon Web Services \\
\small\texttt{wangyida@amazon.com}
\and
Jinyang Li \\
\small New York University \\
\small\texttt{jinyang@cs.nyu.edu}
}

\date{}
\maketitle

\begin{abstract}
Distributed training is essential for scaling the training of large neural network models, such as large language models (LLMs), across thousands of GPUs.  
However, the complexity of distributed training programs makes them particularly prone to \bugs, which do not produce explicit error signals but lead to incorrect training outcomes.
Effectively detecting and localizing such \bugs in distributed training is challenging.
Common debugging practices based on monitoring training loss or gradient norm curves are indirect, inefficient, and provide no way to localize bugs.

To address those challenges, we design and implement \system, the first systematic differential testing system for detecting and localizing \bugs in distributed training.
\system aligns intermediate tensors from distributed training with those from a trusted reference implementation.
To properly compare the floating-point values in the corresponding tensors, we propose a novel mathematical analysis that provides a guideline for setting tolerances, enabling \system to distinguish bug-induced errors from numerical errors.
Experimental results demonstrate that \system effectively detects 11 existing bugs and 3 new bugs in the widely used Megatron-LM framework, while requiring fewer than 10 lines of code changes. \system is effective in various training recipes, including low-precision recipes involving BF16 and FP8. Notably, a popular open-source training framework has already adopted the method proposed by \system in its development workflow.
\end{abstract}

\section{Introduction}
\label{sec:intro}

The field of machine learning (ML) is increasingly driven by scaling to larger models with billions to trillions of parameters~\cite{deepseekv3, llama3}, which have demonstrated growing capabilities in multiple fields~\cite{llm-survey, diffusion-survey}.
This pursuit of scale has transformed modern training pipelines into intricate, resource-intensive systems.
Various distributed training techniques have been proposed to accommodate
the training workload, such as data parallelism~\cite{dp}, tensor parallelism~\cite{megatron}, pipeline parallelism~\cite{pipedream}, and distributed optimizer~\cite{zero}.
Supporting a customizable combination of all these techniques has led to significant implementation complexity in distributed training systems.

These complex distributed training systems introduce new reliability issues in the form of a growing number of \bugs~\cite{uva-egraph, silent-bug}.
Silent bugs~\cite{slapo-bug, te-overlap-bug, cp-dqkv-bug, moe-bug} 
refer to bugs that do not produce crashes, hangs, or other explicit error signals, but lead to an incorrect training outcome that is unexpected by the system developer.
Such \bugs may stem from wrong data~\cite{slapo-bug, cp-dqkv-bug}, or 
incorrect communication operations for cross-device synchronization~\cite{te-overlap-bug, moe-bug} in the training framework.
These bugs may degrade the quality of the trained model, thereby wasting significant training time and massive GPU resources.
Therefore, it is critical to detect and localize silent bugs in distributed training programs before the actual training commences.

\begin{figure}[htbp]
    \centering
    \includegraphics[width=0.85\linewidth]{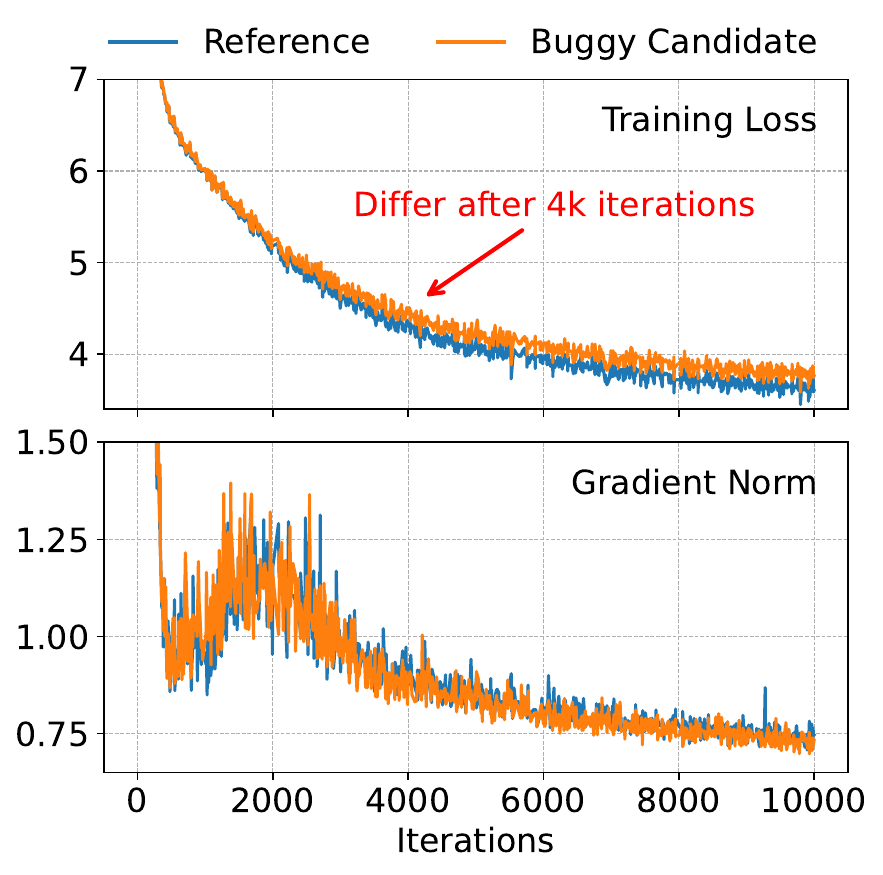}
    \vspace{-\intextsep}
    \caption{The loss and gradient norm curve of a correct and buggy training. The bug can only be observed after 4k iterations from training loss curve.}
    \label{fig:teaser}
    \vspace{-\intextsep}
\end{figure}

Given a distributed training program, how do we tell whether it is correct?
One attractive approach, which is widely adopted in the industry practice, is to do differential testing by comparing the candidate distributed implementation with a trusted reference implementation.
Specifically, developers rely on the alignment of loss and gradient norm curve patterns between the reference and candidate runs on a smaller scale model.
However, 
as illustrated in \Cref{fig:teaser}, it can take over 4,000 iterations (more than 6 hours) for two curves to show a 3\% discrepancy.  Furthermore, when the potential issue is detected in this way, the developer still faces the onerous task of manually locating the whereabouts of the \bug.  Recent concurrent works explore verification-based solutions to check the correctness of distributed training, utilizing SMT solvers or graph rewriting on top of graph-based execution plans~\cite{trainverify,uva-egraph,graphguard}, but they introduce additional deployment challenges (\Cref{sec:background:verification}).

Ideally, we want a systematic differential testing solution that can 
determine if the distributed implementation matches the reference implementation quickly with much fewer resources, e.g., by running a single iteration of stochastic gradient descent (SGD).  Furthermore, the testing should be done in a fine-grained way to compare intermediate tensors during the iteration so that one can pinpoint the whereabouts of potential silent bugs.

Developing such a solution faces two primary challenges.
The first challenge is to align the tensors between the candidate and reference implementations. 
Distributed training employs strategies such as tensor and pipeline parallelism to partition data, models, and intermediate results across devices.
These sharded tensors can be reordered and physically fragmented in multiple ways, making it difficult to find the correspondence between the candidate implementation and the reference implementation.
Secondly, it is non-trivial to determine whether the candidate implementation matches the reference based on the numerical tensor values. 
As floating-point arithmetic is non-associate due to numerical round-off errors, mathematically equivalent operations with different computation orders in the candidate implementation can generate different numerical results as the reference. 
And these numerical errors can further accumulate over multiple operations in the deep neural network (DNN) computation.
This issue is exacerbated by modern training recipes that employ low-precision floating-point representations (e.g., BF16, FP8). 
Therefore, estimating these expected numerical errors and distinguishing them from actual bug-induced errors are critical, as strict bit-wise comparison is impossible and using an ad-hoc threshold is unreliable.

In this work, we propose \system, a streamlined and user-friendly differential testing system to detect and localize \bugs in distributed training. 
To compare a candidate against a reference implementation, in addition to the common practice of initializing model parameters and input data with the same random values, \system also ensures: 
1) intermediate tensors are automatically collected and aligned with minimal user annotations, enabling bug detection and localization on the PyTorch module granularity (addressing challenge \#1);
2) intermediate tensor values are compared using novel and reliable criteria based on input perturbations to the reference model (addressing challenge \#2).

Developers working with the widely adopted distributed training framework Megatron-LM can easily integrate \system with the framework, requiring \textbf{fewer than 10 lines of Python code}, along with annotations specifying the sharding dimensions of the model and the intended distributed training strategy.
In addition to framework users, framework maintainers responsible for testing pipelines can make use of \system and its methodology to guard against changes that may affect numerical stability and correctness. 
Notably, the maintainers of one of the leading open-source frameworks have already adopted this methodology.

In summary, this work makes the following contributions:
\begin{itemize}[leftmargin=*]
\item We identify the \bug as a pain point in distributed training by collecting and reproducing a taxonomy of real \bugs from the widely used Megatron-LM framework.
\item We implement \system, to the best of our knowledge, \emph{the first} system that enables detection and localization of \bugs in diverse practical training settings, including the emerging FP8 precision.
\item We proposed a novel method to systematically estimate the threshold to differentiate the numerical errors from bug-induced errors in distributed training. 
\item We conducted comprehensive evaluations and validated the effectiveness of \system by finding 
11 existing \bugs and 3 new \bugs in Megatron-LM.
\end{itemize}

\section{Motivation, Challenges and Our Approach}
\label{sec:background_and_challenge}

In this section, we will first discuss the current industry practice to debug distributed training based on ad-hoc differential testing and its limitations.  We then examine recent work proposing verification-based methods and explain why they are harder for adoption than differential testing.  Next, we explain the challenges of performing systematic differential testing and briefly discuss how we address them.

\subsection{Industry Practice: Ad-Hoc Debugging}
\label{sec:ad-hoc-debugging}

As discussed in \Cref{sec:intro}, the predominant industry practice for training bug detection is based on differential testing that compares a distributed implementation's output to that of a simpler, trusted reference implementation.  However, what to compare as well as the criteria for detection is currently done in an ad-hoc manner.  Specifically, developers pick a few metrics related to training quality for comparison (e.g., loss rate and gradient norm) and rely on subjective judgment to determine whether there is sufficient deviation to indicate bugs.
While intuitive, such ad-hoc debugging is extremely inefficient.  It can take many thousands of iterations for training quality of a buggy implementation to noticeably differ from the reference run (\Cref{fig:teaser}), leading to tremendous time and compute resource waste.

Furthermore, once a bug is suspected, localizing it requires tedious manual work: extracting intermediate tensor values and comparing them with their counterparts in the reference implementation. Any sufficiently large difference is viewed as a clue to narrow the location of the bug. 
First, the developer must determine how the distributed implementation corresponds semantically to the reference version to decide \emph{which intermediate tensors to compare} -- a mapping that varies across framework configurations and model architectures. 
This effort is difficult to reuse or share, since there is no principled way to explicitly record such correspondences. Even after identifying which tensors to compare, extracting their values is nontrivial -- printing, setting breakpoints, or implementing custom comparison logic is complicated by the opacity of AutoGrad systems in modern training frameworks.

Second, after tackling the engineering challenge of obtaining intermediate tensor values, developers face another daunting problem: what are the correct criteria for comparing them to identify a meaningful difference? This question has existed since the introduction of floating-point numbers, but has recently become much more challenging due to the adoption of low-precision data types like BF16~\cite{mix-precision-bf16} and FP8~\cite{fp8,deepseekv3,transformer-engine} in large-scale training.
These formats inherently introduce larger numerical variation between implementations. 
Most developers are aware that comparing tensor values requires some combination of absolute and relative tolerance to avoid false alarms or missed bugs. Yet even experienced developers may struggle when actually implementing these criteria.

\subsection{Challenges in Verification-Based Methods}
\label{sec:background:verification}
The difficulty of debugging for distributed training has motivated several recent verification-based solutions~\cite{trainverify,uva-egraph, graphguard}.
These works, which are concurrent with our own, aim to prove semantic equivalence of the computation graphs generated in the distributed and single-device settings.  One major benefit of these works is that they can produce a proof artifact that formally establishes the
correctness of the computation graph if verification succeeds. Nevertheless, there are challenges in adopting these methods in production.

TrainVerify~\cite{trainverify} symbolizes the graph-based execution plans from the auto-parallel frameworks, then converts them into logical formulas, and verifies their mathematical equivalence using SMT solvers.
Their workflow has several limitations, which may make it challenging to deploy in production for any distributed training environment.
First, implementation and maintenance cost may be high: users of TrainVerify need to implement and maintain a separate code base from the parallelization framework itself to convert the model and its execution plan into a symbolic data flow graph for verification, along with code that interprets symbolic operators by emitting SMT formulas that accurately describe the operator semantics given all possible combinations of parameters; additionally many distributed frameworks (e.g., Megatron-LM) do not provide out-of-the-box the lineage metadata required by TrainVerify to establish the proof targets.
Secondly, the ``parallelization equivalence'' established by TrainVerify on the computation graphs may not actually guarantee desired behavior in training, since it does not rule out bugs that exist on a lower-level than the computation graphs, such as misuse of custom kernels~\cite{cp-dqkv-bug}, or issues due to numerical instability of mathematically equivalence expressions since the verification is based on real variables.
Finally, localizing issues to specific components may be challenging if the SMT solver based approach fails to
prove the equivalence of two expressions.

By contrast, differential testing is intuitive for the developers, and it has several advantages compared to these verification-based solutions.  First, under differential testing, the framework code and kernels being executed in the debugging process is usually the same as what will be executed during the actual training process. This can save the developers from efforts in 
writing and maintaining separate specification required for verification.
Additionally, executing the actual code directly can potentially expose GPU kernel-related issues
and other low-level problems that are abstracted away in order for verification to be tractable.
Second, once a bug is successfully reproduced and fixed by differential testing, the developers can usually
transform their debugging code to a test in an automated pipeline to prevent regression.

\subsection{How \system Addresses the Challenges of Ad-Hoc Debugging?}
\system extends the ad-hoc debugging procedure described in \Cref{sec:ad-hoc-debugging} in a systematic way with its challenges addressed.
Here we give an overview for how \system addresses the two main challenges:
1) In order to compare intermediate tensors between the two programs, how to map each tensor in the distributed training to its counterpart in the reference implementation?
2) As training involves floating point operations, 
how to distinguish expected numerical errors from bug-induced errors?

\subsubsection{Challenge \#1: Mapping of Semantics}
\label{sec:challenge:mapping}

To align the distributed candidate with the reference, 
we need to build up the mapping of neural network layers and 
the tensor shards in distributed training
to their reference counterparts. 
Building such mapping is challenging for two reasons.
On one hand, different layers or data batches in the model can be partitioned across GPUs using techniques like pipeline parallelism (PP) and data parallelism (DP),
so we may need to collect and rearrange the tensors to 
restore the correct ordering of layers and batches in the single-device reference implementation. 

On the other hand, tensors can also be partitioned in the distributed implementation
under techniques like tensor parallelism~\cite{megatron} (TP), 
sequence parallelism~\cite{sequence-parallelism} (SP), and context parallelism~\cite{ring-attn} (CP).
\system assembles tensor shards according to the particular parallelization strategies to compare tensors.
It achieves this by building a \textbf{tensor canonical mapping} system to set up the alignment.
The mapping system uses a user-provided, intuitive annotation scheme for developers to specify the semantics of the distributed training framework
to construct such tensor canonical mapping, 
as will later be illustrated in the step-by-step description in \Cref{sec:overview}.

\subsubsection{Challenge \#2: ``Expected'' Numerical Errors}
\label{sec:2nd-challenge}
The second challenge is to
\emph{differentiate numerical errors from bug-induced errors}. 
Existing solutions sidestep this by casting all tensors to higher precisions like FP32 or converting all computations to operate on finite fields~\cite{wu_multi-level_2024}, thereby reducing numerical errors to make bug-induced errors more apparent.
Implementing these methods in training frameworks can be difficult, since
widely-used kernels like FlashAttention~\cite{flash-attn} have hard-coded optimizations that relies on BF16 and FP8 data types.

To address this challenge in a way that is non-intrusive to the framework internals and lower-level implementations, \system proposes an empirical numerical error tolerance estimation procedure to distinguish bug-induced errors from numerical errors.
The procedure is backed by theoretical derivations on the expected growth of numerical errors as more neural network layers
are stacked together, based on the assumptions that these layers are locally smooth.
Under this assumption, \system uses a \textbf{perturbation-based FP error estimation} technique to estimate
the ``expected'' accumulation of numerical errors.
Our criteria for comparing tensors uses relative differences of Frobenius norms of tensors~\cite{anla},
i.e., $\text{rel\_err}(A, B)=\|A-B\|/\|A\|$.
The norm is used since it can be calculated regardless the tensor dimension and is less sensitive to individual entries.
When sampling the perturbation in the error estimation procedure, the notion of machine epsilon ($\varepsilon_{\mch}$) or machine precision
is also used: for a datatype, the machine epsilon refers to the relative error of the round-off error (difference between a number and its representation in that 
floating-point format). Intuitively, to simulate numerical errors for different data types, perturbations with
different magnitudes should be used according to the precisions of those data types.  We detail the technique in \Cref{sec:design:err-est} and \Cref{sec:theory}.

\section{Overview of \system Workflow}
\label{sec:overview}
\begin{figure}[!ht]
    \includegraphics[width=0.98\linewidth]{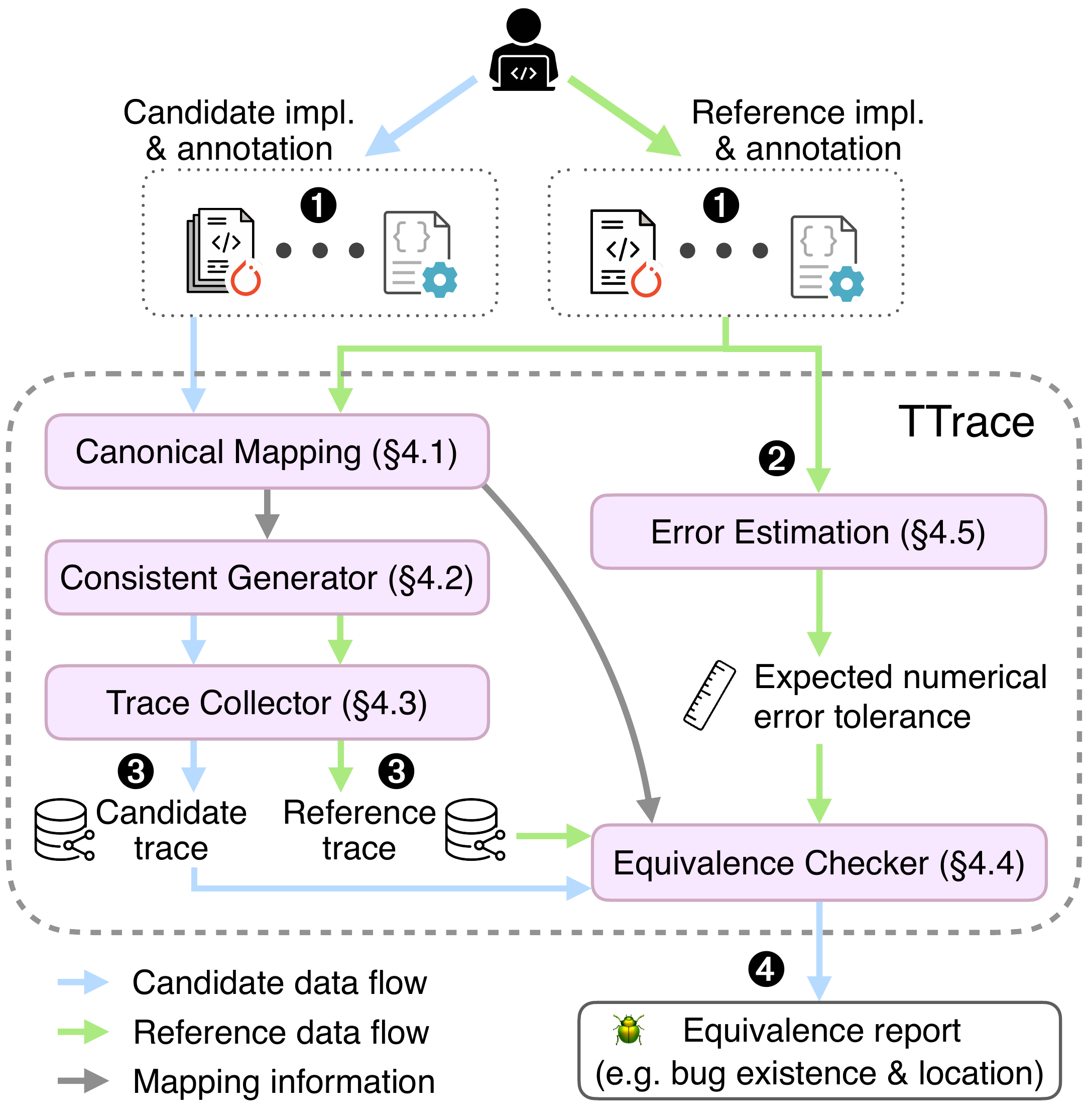}
    \vspace{-0.7\intextsep}
    \caption{The system overview of \system.}
    \label{fig:system-overview}
\end{figure}

\Cref{fig:system-overview} illustrates the system overview of \system. In this section, we describe how various components of \system work together to detect and localize a \bug step by step.

\begin{figure}[!ht]
\centering
\includegraphics[width=\linewidth]{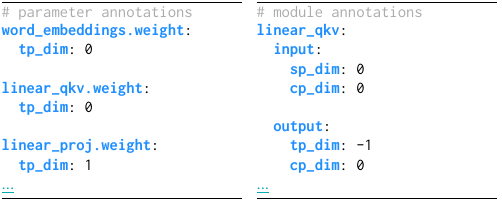}
\vspace{-1.8\intextsep}
\caption{Example annotation clips for a GPT model. 
Module annotations specify which input/output tensors to trace.
The numbers specify the tensor dimensions being partitioned by the particular parallelism.}
\label{lst:annotation}
\vspace{-\intextsep}
\end{figure}

\paragraph{Step 1. User annotates the model.} 

This step identifies which tensors \system should trace and how they are partitioned.
As shown in \Cref{lst:annotation}, 
\emph{parameter annotations} (left) describe how weights are partitioned, such as the word embeddings being sharded along the first dimension.
Simultaneously, \emph{module annotations} (right) specify the target input/output activation tensors to trace and their respective parallel dimensions.
\system uses these specifications to compute a \textbf{tensor canonical mapping} (\Cref{sec:design:canonical-mapping}), allowing it to align and compare tensors between the reference and candidate implementations.

\begin{figure}
\centering
\includegraphics[width=0.9\linewidth]{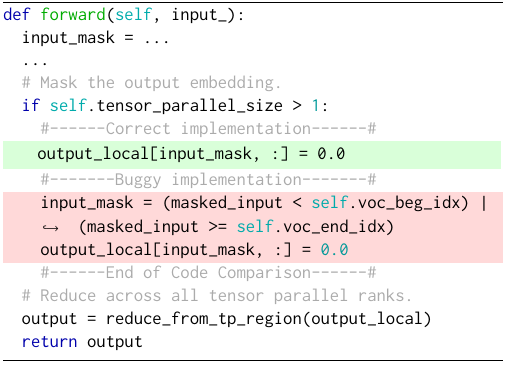}
\vspace{-0.8\intextsep}
\caption{The root cause of the bug in the case study.}
\label{lst:embedding-bug}
\end{figure}

\paragraph{Step 2. \system estimates the expected numerical errors in the model.}

To distinguish numerical errors from bug-induced errors, \system estimates the tolerance of expected numerical errors for the given model using the reference implementation (\Cref{sec:design:err-est}).
The tolerance is later used to compare candidate and reference tensors in the \textbf{equivalence checker} (\Cref{sec:design:equiv-checker}).

\paragraph{Step 3. \system runs candidate and reference.}

\system executes the reference and candidate implementations for a single training iteration, repspectively.
It employs a \textbf{consistent distributed tensor generator} (\Cref{sec:design:consistent-generator}) to ensure identical inputs and a \textbf{trace collector} (\Cref{sec:design:trace-collector}) to capture target activations and gradients during the forward and backward propagation.

\paragraph{Step 4. \system conducts differential testing to detect bugs.}

\system compares the reference and candidate traces.
It generates a report quantifying the discrepancies, and marks values that exceed the error tolerance computed in step 2.
This allows the users to pinpoint exactly which tensors in the candidate implementation diverge from the reference.

\paragraph{Step 5. \system rewrites module inputs to localize bugs.}

A bug in an early layer often leads to cascading errors, causing all subsequent tensors to diverge from the reference (\Cref{fig:fp-bug}a).
To isolate the root cause, \system performs \emph{module-wise testing} (\Cref{sec:design:trace-collector}) by rewriting input tensors.
\system utilizes the \textbf{consistent distributed tensor generator} (\Cref{sec:design:consistent-generator}) to rewrite inputs for every module during execution, ensuring the reference and candidate to consume identical inputs.
This decoupling prevents error from propagating further and eliminates accumulated numerical deviations.
For example, \Cref{lst:embedding-bug}, the root cause for \Cref{fig:fp-bug}a, is a tensor parallelism bug in the word embedding module. It normally cascades through the model. While with module-wise testing, \system curbs the discrepancy to the faulty module, allowing for precise localization.

\section{System Design}
\label{sec:design}

\subsection{Canonical Mapping}
\label{sec:design:canonical-mapping}
\begin{figure}[!ht]
    \centering
    \includegraphics[width=0.8\linewidth]{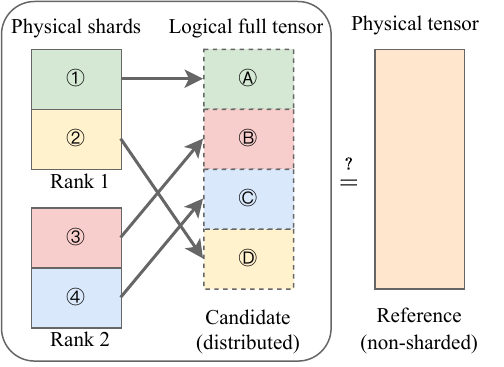}
    \caption{Illustration for canonical shard mapping.}
    \label{fig:shard-mapping}
\end{figure}

To compare tensors in the candidate and reference implementations,
it is essential to establish a rigorous correspondence between the tensors.
The notion of a \emph{logical full tensor} in the candidate (distributed) runs is useful for the discussion.
Such a tensor is assembled from multiple physical shards on multiple devices, and
is expected to match a physical tensor in the non-sharded reference trace (see \Cref{fig:shard-mapping}).
We therefore need to solve two problems with the mapping:
1) how to uniquely identify a tensor in the reference trace or a logical full tensor in the candidate trace; and
2) how to map the local tensor shards in the candidate trace to their corresponding logical full tensors.

\begin{figure}[!ht]
\hspace{-10pt}
    \centering
    \includegraphics[width=0.9\linewidth]{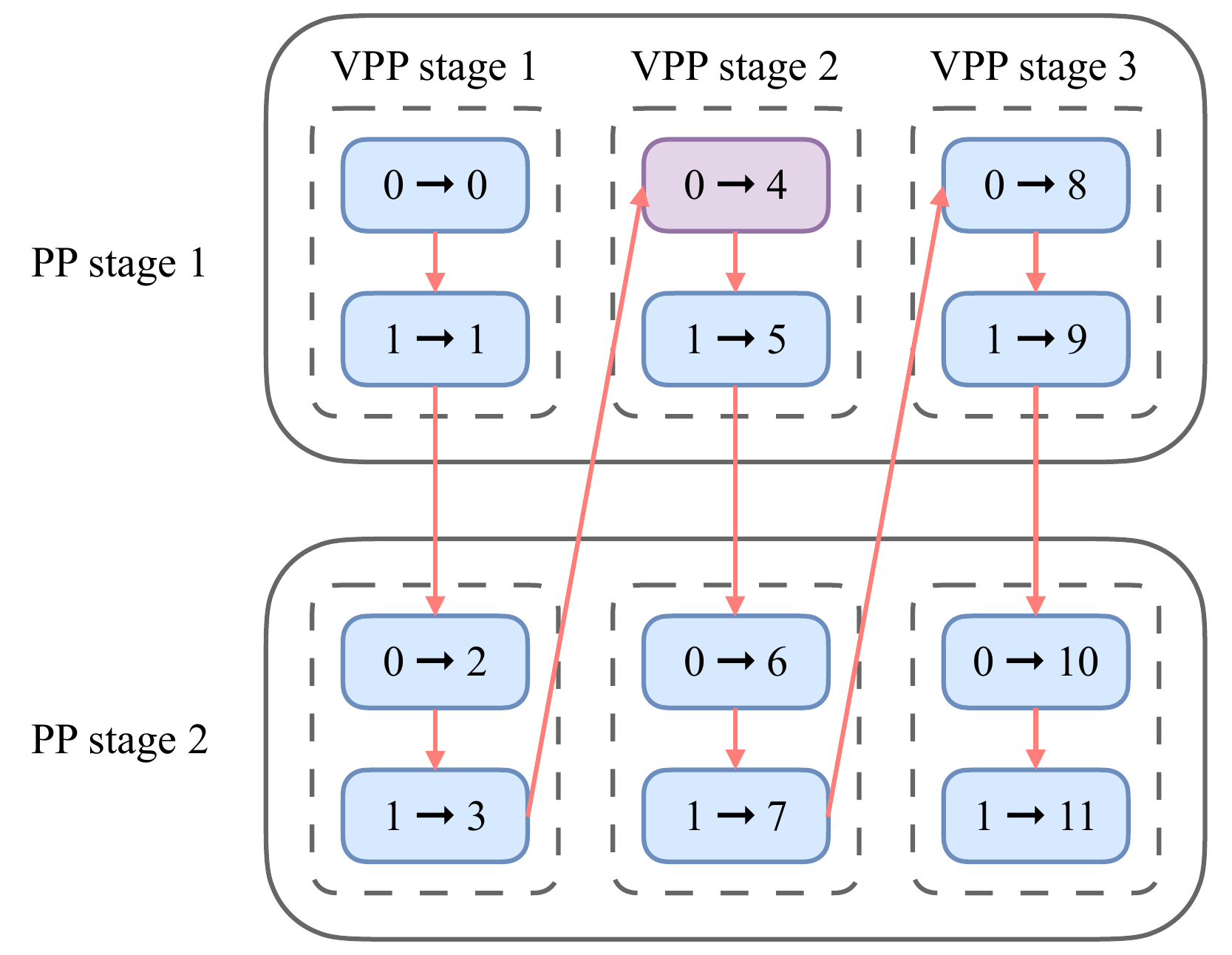}
    \caption{Illustration of the layer index mapping in pipeline parallelism and virtual pipeline parallelism for canonical module name. Each colored box is one layer. The purple example means layer 0 in the 2nd virtual pipeline of the 1st pipeline stage maps to layer 4 in the reference.}
    \label{fig:pp-mapping}
\end{figure}

For the first problem, we introduce a \emph{canonical identifier} to uniquely identify tensors under non-sharding parallelism strategies, such as DP and PP,
distinguishing logical full tensors that are physically separated by data batches or pipeline stages.
This identifier is a composite key consisting of the iteration number, micro-batch index, tensor type, 
and the canonical module name.
Each component targets a specific parallel dimension.
Specifically, the micro-batch index aligns parallel data batches in the candidate under DP with sequential gradient accumulation steps in the reference.
Meanwhile, the canonical module name handles PP by mapping stage-local layer indices (which typically reset to 0) back to global layer indices in the model, as illustrated in \Cref{fig:pp-mapping}.

For the second problem, 
we maintain a data structure called \emph{canonical shard mapping} to 
handle the parallelism strategies that shard the tensors.
It record the mapping from the slices in local shards to those in the corresponding logical full tensor,
as shown in \Cref{fig:shard-mapping}.
A tensor shard on one device can consist of multiple non-contiguous slices in the logical full tensor,
e.g., striped attention in context parallelism.
In this example, we use the shards from the two ranks to compose the full tensor according to the mapping,
and then compare with the physical full tensor from the reference run.
When merging the shards into the full tensor, \system will check to ensure that there is no overlap or omission.

Unlike auto-parallelization systems including nnScaler~\cite{nnScaler} where the tensor mapping are generated when making the parallelization plan, common manually parallelized frameworks like Megatron-LM lacks such mapping information.
\system uses the user-provided tensor annotations to infer the canonical shard mapping.
For example, in \Cref{lst:annotation}, the annotation for the parameters and the input or output activations
contain the dimensions on which particular parallelism strategies split.
The mapped slice is inferred from the sharding dimension, parallel rank, and the semantic of the framework.

Developers can extend \system with new parallelization strategies by providing a function that maps the parallelization ranks to shard mappings.
In practice, this process requires minimal implementation overhead; existing support for parallelism strategies (e.g., PP, TP, SP, CP) typically comprises approximately 30 lines of code each.

For large models where a single-device reference implementation is not available, we run the reference implementation with pipeline parallelism.
Similar to the candidate implementation, the canonical mappings can be established for the reference to facilitate precise comparison with the candidate.
More details regarding the rationale and scalability of the reference implementation can be found in \Cref{sec:eval:scalability}.

\subsection{Consistent Tensor Generator}
\label{sec:design:consistent-generator}
As demonstrated in \Cref{sec:overview}, we need to generate 
the \emph{same} tensor as input to the candidate and reference modules in order to compare their behavior to identify bugs,
as well as when localizing bugs in candidates with the rewriting functionality.
These inputs include the model parameters, the input data to the neural networks, and the rewritten intermediate tensors.
If we randomly generate a tensor for the reference, 
we have to ensure that the same logical full tensor is generated for the candidate, with each rank holding its corresponding shard.
To achieve this, we use the hash of the tensor's canonical identifier as the seed for the random number generator.
This produces the same tensor for the reference implementation and the corresponding logical full tensor for the candidate.
The actual distributed tensors supplied to the candidate are then extracted from 
the generated logical full tensor as slices or shards,
based on the shape and slice information from the canonical mapping in \Cref{sec:design:canonical-mapping}.
In addition to generating consistent input tensors to the modules,
this mechanism can also be used to generate consistent main gradients
to examine the optimizer behavior in the candidate and reference implementations.

\subsection{Trace Collector and Tensor Rewrites}
\label{sec:design:trace-collector}

The trace collector interfaces with the training framework and the underlying deep learning framework
to intercept and record tensor values for comparison.
We prioritize a non-intrusive design with minimal user code modifications leveraging PyTorch’s native APIs:
activations and their gradients are collected through module hooks; parameter gradients are collected through tensor hooks; the main gradients (high-precision accumulator in mixed-precision training) and parameters are collected before and after the optimizer step.

Beyond raw tensor values, canonical identifiers, and shard mappings, 
we also record auxiliary metadata in the trace, 
such as the name and class of the module producing the tensor,
to facilitate the checking and bug localization process.
The traces are stored in the main memory and persisted to disk when the training ends.

As introduced in Step 5 (\Cref{sec:overview}), the tensor rewrite functionality extends this infrastructure to enable active bug localization via module-wise testing.
It utilizes the same underlying hook-based instrumentation architecture as the trace collector.
However, instead of recording tensors, it overwrites the inputs to each module with tensors produced by the consistent distributed tensor generator (\Cref{sec:design:consistent-generator}), without altering the computation graph.
This ensures that corresponding modules in the reference and candidate implementations consume identical inputs.
Consequently, this mechanism isolates the computation of each module, preventing the errors from propagating to downstream computations, thereby enabling precise bug localization.

\begin{figure}
    \centering
    \includegraphics[width=0.8\linewidth]{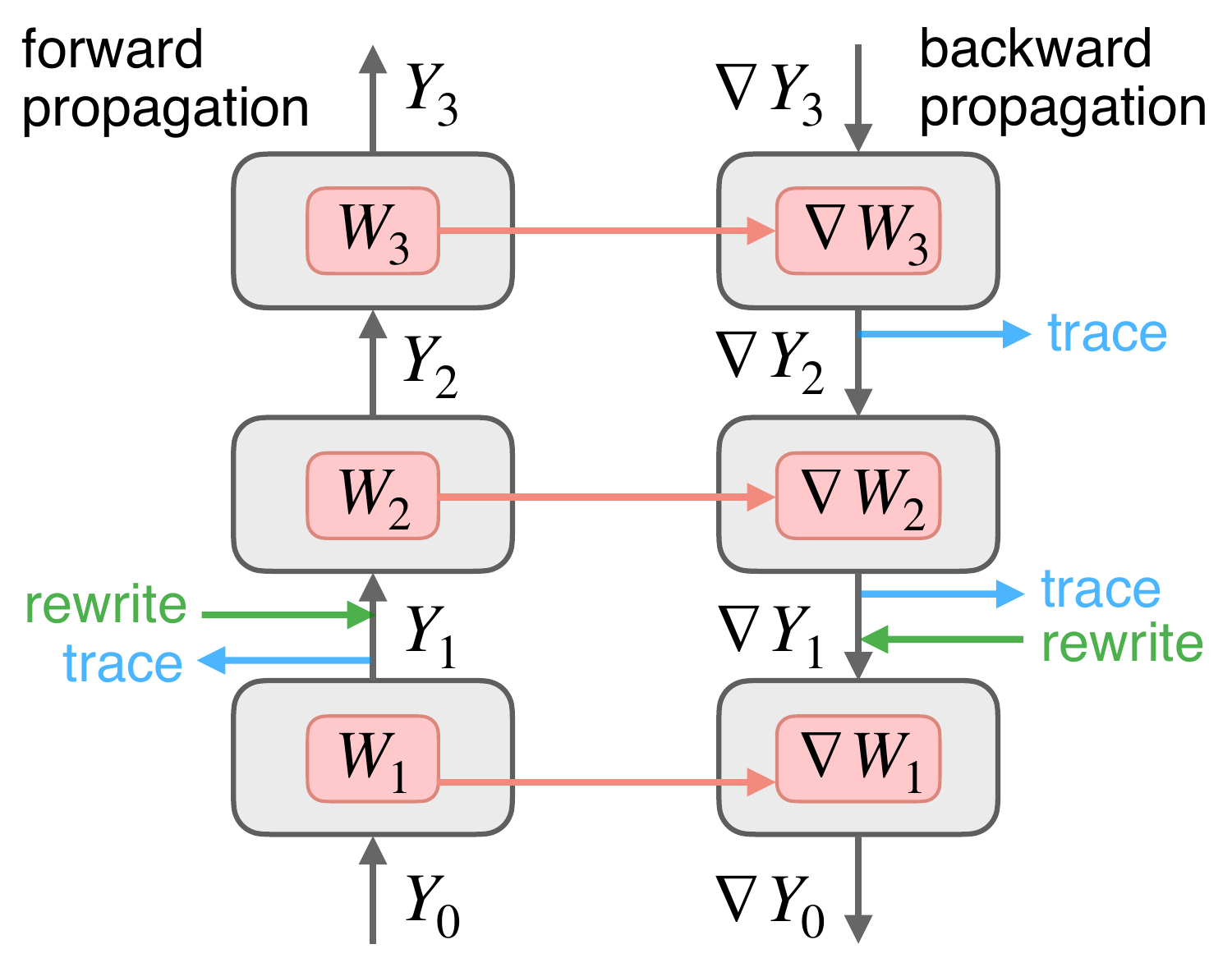}
    \caption{Illustration for trace collector and tensor rewrites in forward and backward propagation}
    \label{fig:placeholder}
\end{figure}

\subsection{Equivalence Checker}
\label{sec:design:equiv-checker}

After capturing the traces from both the candidate and reference executions, \system uses the equivalence checker to determine whether 
the two traces are equivalent.
The process involves two stages.
First, it aggregates the distributed tensor shards to reconstruct the corresponding logical full tensors, 
enabling direct comparison with the reference tensors.
Second, it performs differential testing on the merged tensors and compares the error between the reference and candidate against the dynamic numerical error tolerances obtained in \Cref{sec:design:err-est}.
% }

\paragraph{Tensor merger and consistency check.}
The merger utilizes the canonical shard mapping (\Cref{sec:design:canonical-mapping})
stored in the traces to reconstruct the logical full tensors from the distributed tensor shards.
Crucially, this process also enforces consistency among the tensor shards and reports potential bugs.
For example, tensor shards that are replicated across devices must hold equivalent values, e.g., in standard data parallel, the main gradients from each rank must be same.
If the merger detects a discrepancy among replicas, such as divergent gradient caused by a missing \texttt{all-reduce}, it immediately flags the specific tensor and reports the bug with its location.

\paragraph{Differential testing.}
Given the merged traces from the reference and the candidate, 
the equivalence checker performs differential testing on them.
First, it computes the differences  between the reference and the candidate.
It then compares the differences against the expected numerical error tolerance obtained in step 2 in \Cref{sec:overview}.
If the difference exceeds the tolerance threshold, the checker identifies the deviation as a bug, and localizes the bug by the location of the particular tensor where the violation occurred.

\subsection{Perturbation-Based Error Estimation}
\label{sec:design:err-est}

\system proposes a novel method of using perturbation responses of models to estimate expected numerical errors at various program locations. 
The estimated errors are provided to the equivalence checker (\Cref{sec:design:equiv-checker}) that compares tensors,
in order to reliably distinguish benign numerical deviations from actual implementation bugs,
without relying on static tolerances.
Let $\hat{F_L}(X)$ denote the numerical implementation of a $L$-layer neural network function $F$ with input tensor $X$, and let $\hat{H_L}$ represent the distributed candidate implementation under test.
During the estimation, we sample a small perturbation $\Delta X$ at random to the input $X$ with $\|\Delta X\|=\Order(\varepsilon_{\mch})$.  We then measure the induced difference between $\hat{F_L}(X+\Delta X)$ and $\hat{F_L}(X)$ as $L$ changes to serve as a guideline for what the expected numerical errors look like. 
For a detailed theoretical justification of why this perturbation-based method effectively estimates numerical errors, please refer to \Cref{sec:theory:estimate}.

During the checking phase, the differential testing calculates the actual difference between $\hat{F_L}(X)$ and $\hat{H_L}(X)$, and compares this observed difference against the estimation. We conclude that $\hat{H_L}$ probably contains bugs if the observed difference between $\hat{H_L}(X)$ and $\hat{F_L}(X)$ significantly exceeds or deviates in pattern from the perturbation difference between $\hat{F_L}(X+\Delta X)$ and $\hat{F_L}(X)$ in the reference implementation.
An empirical evaluation of this strategy across popular Transformer architectures is detailed in Section~\ref{sec:eval}.

\section{Error Tolerance Analysis}
\label{sec:theory}

To effectively detect silent bugs, one must distinguish between \emph{expected tensor differences} (inherent to FP semantics) from differences caused by logical errors.
However, analytically tracking the accumulation of internal round-off errors to compare a distributed implementation against a reference is computationally intractable.
It can also be shown that the familiar method of setting through experience absolute/relative tolerances as estimations of the expected errors can be highly inaccurate (\Cref{sec:eval:fp-err}).
To address this, we introduce a \textbf{perturbation-based proxy} to estimate these errors.
Our core insight is that a network's sensitivity to internal round-off errors is structurally equivalent to its sensitivity to input perturbations. 
Since both error sources propagate through the same computation graph, they are governed by the same layer-wise smoothness conditions.

In this section, we provide the theoretical grounding for this proxy. 
We first estimate error bounds within the forward and backward propagation (\Cref{sec:theory:bounds}) and prove that Pre-LayerNorm Transformer layers are likely well-conditioned at random initialization. 
We then justify why the input perturbation based method described in \Cref{sec:design:err-est} can serve as a reliable criteria for detecting bugs (\Cref{sec:theory:estimate}).

\subsection{Analysis of Expected Numerical Errors}
\label{sec:theory:bounds}
For this analysis, we denote the underlying function that a neural network is computing as a function on real-valued tensors $Y = f_l(X)$. 
Let $F_L$ denote the composition of $L$ layers: $F_L = f_L \circ f_{L-1} \circ \dots \circ f_1$.
We recursively define the intermediate tensors as $Y_0 = X$, $Y_l = f_l (X_{l-1})$. 
Correspondingly, we let $\hat{f_l}$ and $\hat{F_L}$ denote the FP implementation of $f_l$ and $F_L$, respectively, with $\hat{Y_l}$ representing the FP tensor corresponding to $Y_l$.

Our analysis relies on the smoothness properties of the neural network layer functions.
We assume each $f_l$ is locally Lipschitz continuous upon initialization. 
For the specific case of Pre-LN Transformers, we derive a bound on the layer-wise Lipschitz constant resulted from random initialization.

\begin{theorem}[Smoothness of Transformer Layers at Initialization]
\label{thm:smooth}
    A Pre-LN transformer layer $f_l$ is well-conditioned with high probability under standard initialization: its Lipschitz constant scales as $C_l \approx 1 + \Order(l^{-\frac{1}{2}})$.
\end{theorem}

Regarding the FP implementation $\hat{f_l}$, we assume that the implementations, whether distributed or not, are stable. Formally, we assume the local relative error is on the order of machine epsilon.
Based on the above assumptions regarding smoothness and FP error characteristics, we derive the following bound on the accumulated error of activations:

\begin{theorem}[Expected Activation Difference Bound]
\label{thm:text_expected_diff_bound}
Let $F = f_L \circ \dots \circ f_1$ be the actual underlying function the neural network is computing,
and $\hat{F} = \hat{f}_L \circ \dots \circ \hat{f}_1$ be a reasonable FP implementation of $F$, where ${Y}_l$ and $\hat{{Y}}_l$ are their intermediate activations. 
A probabilistic bound on the expected relative difference of the difference norm when comparing the final outputs is:
\vspace{-5pt}
\[
\E\left[ \frac{\norm{\hat{{Y}}_l - {Y}_l}}{\norm{{Y}_l}} \right] \lessapprox \Order \left( \eta \cdot \sqrt{l} \cdot \varepsilon_{\mch} \right) 
\]
where $\eta$ is a constant related to numerical stability of the FP implementation of each layer $\hat{f_l}$.
\end{theorem}

We prove another theorem on the expected relative difference of gradient norms in transformers as follows.
Future work is needed to generalize this analysis to other neural network architectures.
\begin{theorem}[Expected Gradient Difference Bound]
\label{thm:text_gradient_diff_bound}
Let $G_{{W}^l}$ and $\hat{G}_{{W}^l}$ be the ideal and FP gradients for a parameter matrix ${W}^l$ in layer $l$ of
a transformer with $L$ layers in total. 
Given the additional assumption that the backward propagation is numerically stable, the expected relative difference norm of the gradient is bounded by
\vspace{-5pt}
\[ \E \left[ \frac{\| \hat{G}^{{W}^l} - G^{{W}^l} \|}{\|  G^{{W}^l} \|} \right] \lessapprox \Order\left( \sigma \cdot \sqrt{\frac{L}{l}} \cdot \varepsilon_{\mch} \right)  \]
where $\sigma$ is a constant related to the numerical stability of the 
back propagation implementation of each layer.

\end{theorem}

\newcommand{\LM}{https://github.com/NVIDIA/Megatron-LM}
\newcommand{\TE}{https://github.com/NVIDIA/TransformerEngine}
\begin{table*}[!ht]
\hypersetup{
    colorlinks=true,        % color text but not frame
    linkcolor=blue!70!black,
    urlcolor=blue!70!black,
    citecolor=blue!70!black,
    pdfborderstyle={/S/U/W 1}  % underline style（Style=Underline, Width=1）
}
\caption{List of \bugs discovered by \system.  The first 11 are existing bugs collected from the issue list of Megatron-LM and TransformerEngine.  The last 3 are found by \system when doing sweep tests for the newest releases of them.}
\vspace{-\intextsep}
\label{tab:bugs}
\begin{tabular}{cclllll}
\toprule
ID& New  & Type & Description                            & Impact                   & Issue                        & Fix                                       \\
\midrule
1 & \no  & WD & TP: wrong embedding mask               & Wrong forward, gradients & N/A           & \href{https://github.com/awslabs/slapo/pull/80}{pull/80} \\
2 & \no  & WD & AR: wrong input                        & Wrong gradients          & \href{\LM/issues/371}{371}   & \href{\LM/commit/fc2c81d}{\code{fc2c81d}} \\
3 & \no  & WD & CP: wrong loss scaling                 & Wrong gradients          & \href{\LM/issues/673}{673}   & \href{\LM/commit/0c2cd58}{\code{0c2cd58}} \\
4 & \no  & WD & DP: wrong loss scaling                 & Wrong gradients          & \href{\LM/issues/906}{906}   & \href{\LM/commit/3bdcbbb}{\code{3bdcbbb}} \\
5 & \no  & WC & ZeRO: embedding and LM-head untied     & Wrong parameter update   & \href{\LM/issues/656}{656}   & \href{\LM/commit/db2040f}{\code{db2040f}} \\
6 & \no  & MC & SP: router weights not synchronized    & Wrong gradients          & \href{\LM/issues/599}{599}   & \href{\LM/pull/619/}{pull/619}            \\
7 & \no  & WC & TP: wrong FP8 communication group      & Wrong forward, gradients & \href{\TE/issues/335}{335}   & \href{\LM/pull/442}{pull/442}             \\
8 & \no  & WD & AR: wrong tensor by FP8 cast           & Wrong loss               & \href{\TE/issues/539}{539}   & \href{\TE/pull/646}{pull/646}             \\
9 & \no  & WC & ZeRO: parameter update failure         & No parameter update      & \href{\LM/issues/663}{663}   & \href{\LM/commit/75120db}{\code{75120db}} \\
10& \no  & WD & PP: wrong stage division               & Wrong model get trained  & \href{\LM/issues/468}{468}   & \href{\LM/commit/8e98a72}{\code{8e98a72}} \\
11& \no  & WC & TP: wrong gradients with overlap       & Wrong gradients          & \href{\TE/issues/1616}{1616} & \href{\TE/pull/1554}{pull/1554}           \\
\midrule
12& \yes & MC & SP: layernorm weights not synchronized & Wrong gradients          & \href{\LM/issues/1446}{1446} & \href{\TE/pull/1528}{pull/1528}           \\
13& \yes & WD & CP: wrong attention gradients          & Wrong gradients          & \href{\TE/issues/1557}{1557} & \href{\TE/pull/1539}{pull/1539}           \\
% 13& \yes & W-CP & TP: wrong attention gradients          & Wrong gradients          & \href{\TE/issues/1615}{1615} & N/A                                       \\
14& \yes & WC & TP+CP: wrong layernorm gradients       & Wrong gradients          & \href{\TE/issues/1677}{1677} & N/A                                       \\
\bottomrule
\end{tabular}

\vspace{.5em}
\begin{minipage}{\linewidth}
\footnotesize
\textbf{WD}: Wrong Data \quad
\textbf{WC}: Wrong Communication \quad
\textbf{MC}: Missing Communication \quad
\textbf{AR}: Activation Recomputation 

\textbf{Issue}: 
The numbers are issue IDs for the bugs; 
N/A means the bug is not reported in an issue.

\textbf{Fix}: 
The hex string means the fixing commit id; 
the numbers are the fixing pull requests; 
N/A means the bug has not been fixed yet.
\end{minipage}
\end{table*}

\subsection{The Input Perturbation Proxy}
\label{sec:theory:estimate}

Complementing the theoretical bounds derived above, we now justify why the input perturbation method (\Cref{sec:design:err-est}) effectively distinguishes bugs from numerical errors.

Let $\hat{F}(X)$ be the reference FP implementation and $\hat{H}(X)$ be the distributed candidate implementation.
We aim to determine if the observed discrepancy $\|\hat{H}(X) - \hat{F}(X)\|$ is permissible.
By the triangle inequality, this discrepancy is bounded by the sum of their individual deviations from the ideal real-valued function $F(X)$:
\vspace{-0.2cm}
\begin{equation*}
\label{eq:triangle}
    \| \hat{H}(X) - \hat{F}(X) \| \leq \underbrace{\| \hat{H}(X) - F(X) \|}_{\text{candidate error}} + \underbrace{\| F(X) - \hat{F}(X) \|}_{\text{reference error}}
\end{equation*}
\vspace{-0.2cm}

If the candidate $\hat{H}$ is a correct and numerically stable FP implementation of $F$,
its forward error must satisfy the same theoretical bounds as the reference implementation derived in \Cref{thm:text_expected_diff_bound}.
Consequently, the permissible difference between the two implementations is effectively on the same order as the expected forward error of the neural network.

Calculating the forward errors relative to the real-valued $F$ exactly is impossible for either the reference or candidate.
However, based on the smoothness property (\Cref{thm:smooth}), the accumulation of the forward error induced by internal round-off is structurally similar to the error induced by an input perturbation of size $\varepsilon_{\mch}$.
Therefore, we use the \emph{perturbation response} of the reference implementation, $\| \hat{F}(X+\Delta X) - \hat{F}(X) \|$, as an empirical proxy to estimate the magnitude of term $\|F(X) - \hat{F}(X) \|$, which is valid due to the local Lipschitz continuous and numerical stability assumptions.
This explains our practical decision rule in \Cref{sec:design:err-est}: if the observed difference $\|\hat{H}(X) - \hat{F}(X)\|$ exceeds the magnitude of the perturbation response $\| \hat{F}(X+\Delta X) - \hat{F}(X) \|$, there is likely a logical error in the candidate implementation $\hat{H}$.

\section{Implementation and Evaluation}
\label{sec:eval}

\system is implemented in about 4000 lines of Python code.
The differential testing module is written in about 200 lines of C++ code to 
bypass the Python global interpreter lock and enable parallel execution.
The implementation is built atop PyTorch, 
leveraging its hook mechanism and AutoGrad system.
To evaluate the practicality and effectiveness of \system,
we focus on the following research questions (RQ).
\begin{itemize}[leftmargin=*]
\item \textbf{RQ1}: Can \system discover \bugs, including previously unknown bugs, in distributed training frameworks? (\Cref{sec:eval:bugs} - \Cref{sec:eval:new-bugs})
\item \textbf{RQ2}: How efficient is \system compared to current debugging practices discussed in \Cref{sec:intro}, and can it scale to large models? (\Cref{sec:eval:scalability})
\item \textbf{RQ3}: Why is dynamic error tolerance estimation necessary? (\Cref{sec:eval:fp-err})
\item \textbf{RQ4}: How accurate is the error tolerance estimation, and can it distinguish bug-induced errors from numerical errors? (\Cref{sec:eval:efficacy})
\end{itemize}

\subsection{Experimental Setup}
We evaluate \system targeting Megatron-LM and TransformerEngine.
The default software versions used in our experiments are:
CUDA: 12.4, cuDNN: 9.8.0, PyTorch: 2.4.1, Megatron-LM: 0.11.0, TransformerEngine: 1.13.0.
Our default hardware testbed is an Amazon EC2 p5en.48xlarge machine with 8 NVIDIA H200 GPUs.
We tested GPT-based language models, including GPT, Llama, and MoE models.

We selected Megatron-LM~\cite{megatron-lm} as our primary target 
because it is one of the most widely-used production-ready training frameworks.
It has been adopted and adapted by major technology companies for large model training, such as NVIDIA~\cite{megatron-lm}, 
Microsoft~\cite{megatron-deepspeed}, Amazon~\cite{sagemaker}, ByteDance~\cite{megascale}.

\subsection{Detected Silent Bugs}
\label{sec:eval:bugs}
\Cref{tab:bugs} shows the \bugs that \system discovered.
We collected the \bugs from the issue page of Megatron-LM and TransformerEngine.
Although most of the \bugs are typically resolved by engineers during the development stage, 
we found 11 \bugs and reproduced them using the corresponding history commits.
\system was able to discover all of them.
The types and impact of these bugs are described in \Cref{tab:bugs}.
These bugs will cause the training to have unintended behaviors and cause consequences like degraded model accuracy.

We also conducted a sweep test on different combinations of 4D parallelism (DP, PP, TP, CP) 
together with the additional techniques of sequence parallelism and virtual pipeline parallelism
on the latest stable releases of Megatron-LM and TransformerEngine.
\system successfully detected 3 new \bugs during the test.
They have all been confirmed by NVIDIA developers, and 2 of them have already been fixed.

\paragraph{False positive and false negative.} 

In our experiments, \system did not raise any false alarms. As theoretically derived in \Cref{sec:theory} and empirically validated in \Cref{sec:eval:fp-err}, 
our method yields accurate estimations of numerical errors under mild assumptions that are realistic in actual model architectures. 
Therefore, \system is unlikely to attribute discrepancies arising from numerical errors to actual bugs, unless the architecture is not
locally smooth or the reference implementation itself is not numerically stable, which can also make the actual training challenging even without bugs.

It is theoretically possible for \system to miss bugs that cause deviations at the magnitude of numerical errors in each SGD iteration, but the accumulated
effect of such deviations may eventually cause training to diverge. \system can also miss bugs that cannot be reliably reproduced when running one iteration of training. 
We have not found reports of these bugs to MegatronLM nor encountered such bugs in our experiments.

\paragraph{Categorization of Silent Bugs}
We refer to the bugs using the ID listed in \Cref{tab:bugs}.
Bug 1 corresponds to the running example discussed in \Cref{sec:overview}.
We categorize the detected \bugs into three types:
\begin{itemize}[leftmargin=*]
\item \textbf{Wrong Data (WD)}:
The input tensor for an operation deviates from what is expected.
This includes the input being stale (bug 2); 
incorrectly scaled (bug 3, 4); 
or in a mismatched layout (bug 13), etc.
\item \textbf{Wrong Communication (WC)}:
The order or pattern of the collective communication is incorrect, 
resulting in an erroneous output tensor.
Examples include an incorrect communication order (bug 5),
misconfigured communication groups (bug 7), 
or a wrong communication pattern when computation-communication overlapping is enabled (bug 9), etc.
\item \textbf{Missing Communication (MC):}
One or more collective communication operations are omitted, leaving the result tensor in an incorrect state.
A representative example is a missing all-reduce, which causes the tensor to remain a partial sum (bug 6, 12).
\end{itemize}

\subsection{Case Studies on New Bugs}
\label{sec:eval:new-bugs}

In this section, we highlight \system's ability to uncover subtle silent bugs in training frameworks (RQ1) through case studies on the new bugs discovered.

Bug 12 results in wrong gradients for normalization layers weights. 
When sequence parallel (SP) is enabled, the normalization layers must conduct an all-reduce across SP ranks to gather the gradients from different sequence partitions.
This reduction should occur explicitly after the forward and backward computations and prior to the weight update. 
However, due to a flag misconfiguration, this all-reduce operation was skipped, leaving the gradient to be incomplete.
This bug is hard to find by other methods because it is a manual step outside of the forward and backward computation graph.
\system identified it by detecting inconsistencies among ranks when merging the main gradients.

Bug 13 corrupts the gradients of self-attention activations when context parallelism is enabled. 
In this case, the forward and backward computations of self-attention are performed using customized kernels.
The bug arose because the implementation failed to ensure memory contiguity for the input tensor before passing it to the kernel, causing the kernel to operate on an invalid memory layout.
This bug is elusive as it does not directly affect the parameter gradients, and it involves calling a custom kernel.
\system pinpointed the root cause to the backward computation implementation by localizing the error to the outputs of the attention gradient module.

Bug 14 generates erroneous parameter gradients for layer normalization when both tensor parallel (TP) and context parallel (CP) are enabled in FP8 training.
The gradients are wrongly scaled by the tensor parallel size due to the use of an improper reduction operator during communication.
Note that this bug is discovered during our comprehensive sweep of parallelism configurations, and it is not present 
when using TP or CP alone, or during BF16 training.
This highlights the potential of \system  to serve as a component in the automated testing pipeline to catch bugs
that occur only with specific configurations.

\subsection{System Efficiency and Scalability}
\label{sec:eval:scalability}
\paragraph{Efficiency}
To address RQ2, we conducted the time cost comparison experiment 
for \system against the common debugging practice mentioned in \Cref{sec:intro}.
We will refer to it as \emph{na\"ive approach} in the following text.
We use bug 1~\cite{slapo-bug} as an example.
The na\"ive approach requires training both the reference and candidate implementation until the loss values diverge significantly.
As shown in \Cref{fig:teaser},
it took more than 4000 iterations for the loss curve to exhibit a 3\% relative error.
However, \system only need to run for one single iteration.
We measured the time spent on training the reference for the corresponding iterations because it is the bottleneck.
It took the na\"ive approach 6 hours and 40 minutes, while \system only needs 54 seconds
to identify the bug.

\paragraph{Large reference models}

If the code path of the candidate implementation is independent of the tensor sizes in the training framework code, correctness of such framework code can be checked with a down-scaled model against a single-device reference implementation.
In our evaluation, all bugs we discovered in \Cref{tab:bugs} can be reproduced and diagnosed using a down-scaled model
whose reference implementation fits on a single device.
However, there are cases where the framework code path may vary depending on the model size.
\system provides a way to run the reference implementation through pipeline parallelism (PP) to support reference models that do not fit in the memory of a single device.

We adopt PP because it is less error-prone and its semantics remain the same as a single-device execution at every pipeline stage. 
Specifically, our PP implementation employs the pipelining functionality provided natively by PyTorch, without additional system optimizations such as activation checkpointing or computation-communication overlap, thereby minimizing potential sources of implementation bugs.
Moreover, the implementation follows the same code path regardless of model size or tensor dimensions, and we validated this PP implementation by cross-checking its outputs with those of a single-device reference on smaller models.  
A further advantage of PP is that it preserves identical numerical computation to the single-device execution, since each module fits within a single GPU's memory and no operator partitioning or computation reordering is introduced.

\begin{table}[ht]
\setlength{\cmidrulewidth}{\lightrulewidth}
\caption{The time for \system reference execution on models beyond single-device capacity.}
\vspace{-0.8\intextsep}
\label{tab:scalability}
\begin{tabular}{ccccccc}
\toprule
Model class & \multicolumn{3}{c}{Llama 3} & \multicolumn{3}{c}{DeepSeek V3} \\ 
\cmidrule(r){1-1} \cmidrule(lr){2-4} \cmidrule(lr){5-7} 
Model size  & 8B      & 70B     & 405B    & 16B      & 236B      & 671B     \\
Time (mins)  & 1.8     & 7.2     & 17.1    & 2.8      & 15.2      & 22.7    \\
\bottomrule
\end{tabular}
\end{table}

\Cref{tab:scalability} shows the tracing times for the pipelined references of models that do not fit on a single device. 
The overheads mostly come from the device to host memory movement.
Notice that the candidate implementations usually run faster than the reference implementations due to optimizations
or more parallelisms enabled.
Even for the DeepSeek V3 671B, one of the largest open source language models, the tracing time of \system remains reasonable.

\begin{figure*}[!ht]
    \centering
    \includegraphics[width=0.95\linewidth]{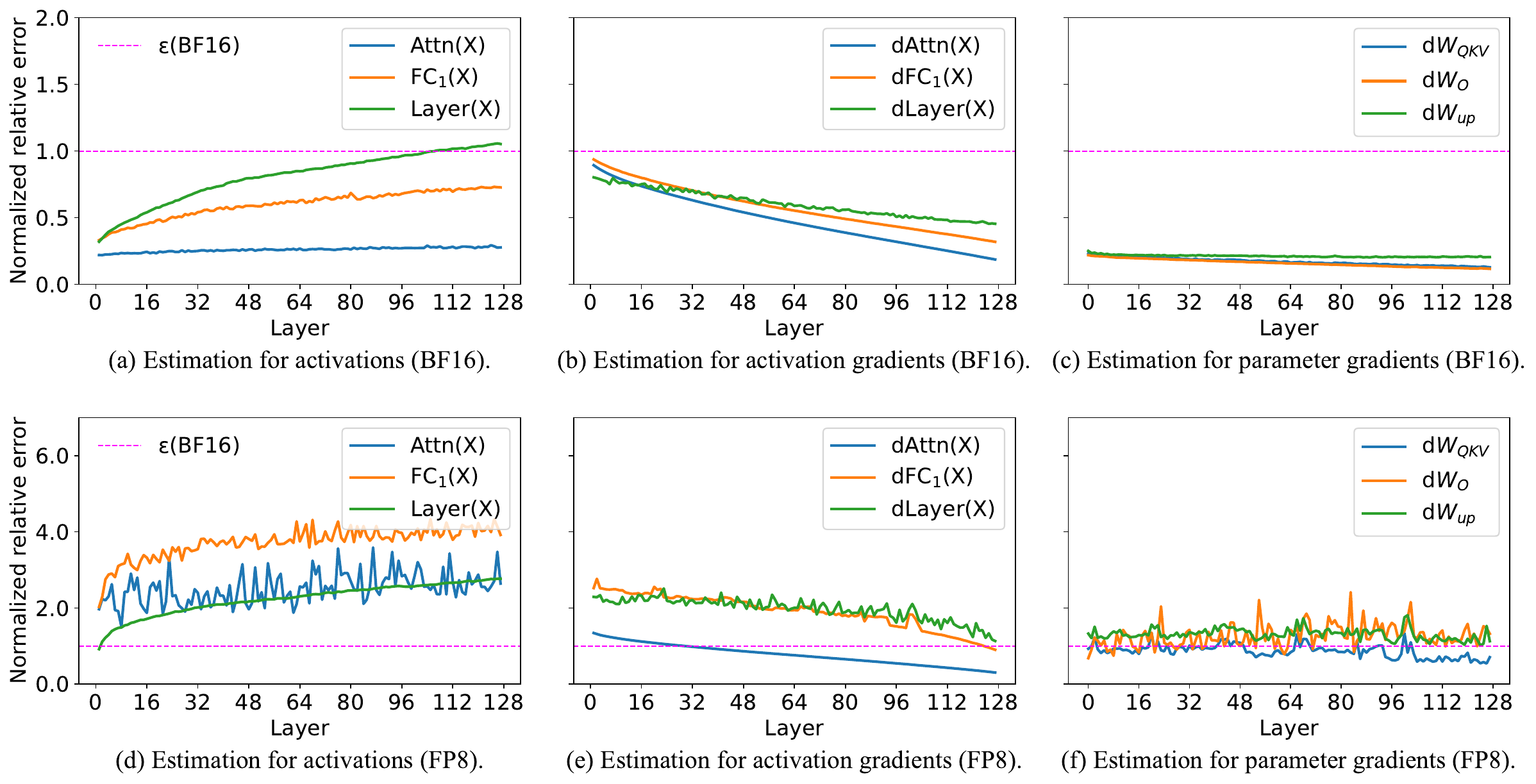}
    \caption{The estimated numeric error tolerance estimation curves regarding layers. We restricted our attention to $L \leq 128$ since most popular transformer architectures have less than 128 layers~\cite{llama3, deepseekv3}, and Transformers can suffer from \emph{rank collapse} if the number of layers is too large \cite{wu_role_2024}.  The y-axis is the relative error normalized by the machine precision of BF16.  The line of machine precision of BF16 is plotted to show the order of magnitude. 
    Note that FP8 curves are normalized by the machine precision of BF16 because the mixed-precision recipe stores the tensors in BF16.}
    \label{fig:fp-bf16}
\end{figure*}

\begin{figure*}[t]
    \centering
    \includegraphics[width=\linewidth]{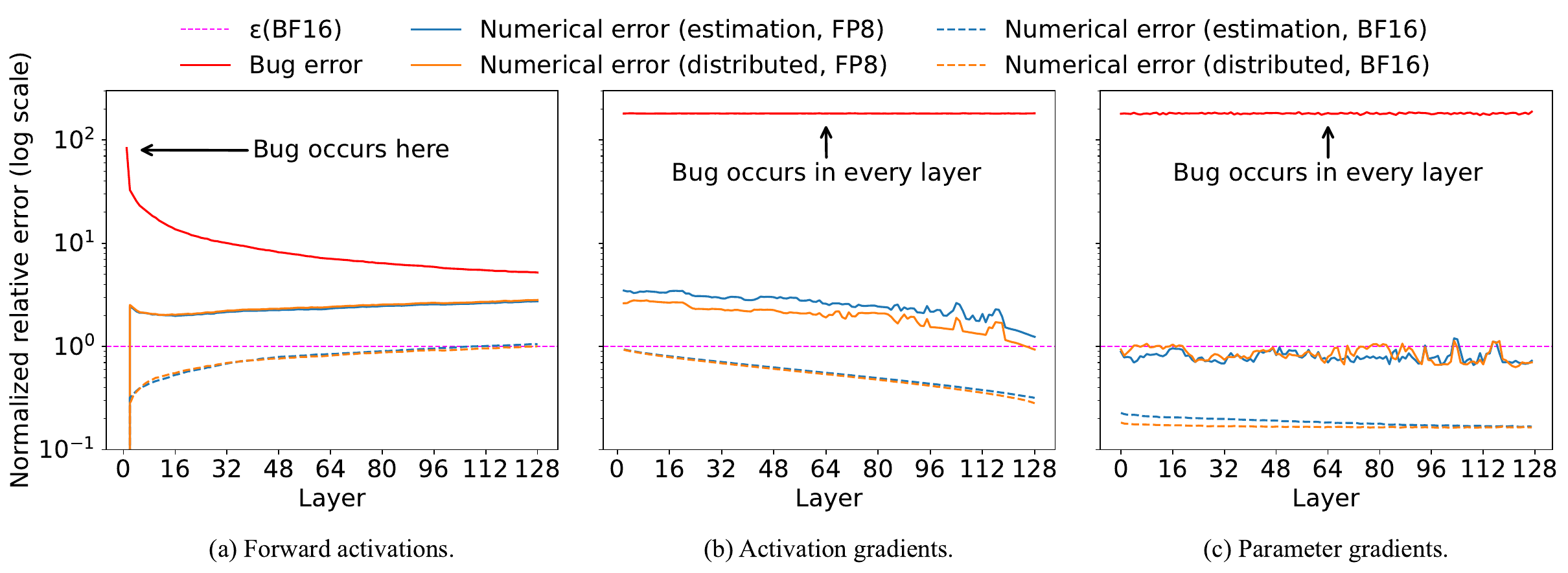}
    \caption{Comparison of bug-induced errors, estimated FP round-off errors obtained through input perturbation, and the actual observed FP round-off errors in the distributed candidate run. The y-axis is normalized by the machine epsilon of BF16. In \Cref{fig:fp-bug}a (forward propagation), we reproduce bug 1, which affects the embedding layer. Consequently, the errors are most significant in the first few layers. In \Cref{fig:fp-bug}b and \ref{fig:fp-bug}c (backward propagation), we reproduce bug 11, which involves incorrect communication. This causes the gradients to be erroneous across every layer. It is clear from this figure
    that bug-induced errors are usually much larger than the numerical errors, even when there is a large number of layers.
    }
    \label{fig:fp-bug}
\end{figure*}

\subsection{Ablations for Dynamic Error Estimation}
\label{sec:eval:fp-err}

In this section, we address RQ3 by demonstrating the necessity of dynamic error tolerance estimation.
We first show in \Cref{fig:fp-bf16} that our estimated numerical errors are influenced by multiple aspects, including the number of layers, data type, and model architecture.
This offers an intuition that our tolerance estimation procedure offers a more fine-grained criteria than a statically determined error tolerance, which may be inappropriate in comparing the intermediate tensors. 
These curves also provide an empirical validation for  \Cref{thm:smooth} that the transformer layers are likely locally smooth at initialization, and also the growth of bounds presented in \Cref{sec:theory}.

We then show in \Cref{tab:tolerance} that setting the correct static tolerances when comparing tensor values is indeed challenging, as different choices either introduce false positives, or false negatives, or both when applying these criteria to the traces \system collected for the bugs in \Cref{tab:bugs}.
We evaluate various combinations of absolute (\texttt{atol}) and relative (\texttt{rtol}) tolerances in \texttt{torch.allclose}, including the PyTorch defaults and looser values popular in practice.
In contrast, no false positive or false negative was observed using the dynamic tolerance estimation from \system.

\begin{table}[ht]
\caption{Comparison of false positives and false negatives for \system's dynamic error tolerance estimation versus the baseline \texttt{torch.allclose} fixed tolerance. The results correspond to the bug set in \Cref{tab:bugs}, `-' denotes no occurrence among these bugs.}
\vspace{-0.8\intextsep}
\label{tab:tolerance}
\begin{tabular}{lcccc}
\toprule
& \texttt{atol} & \texttt{rtol} & \begin{tabular}[c]{@{}c@{}}False\\ positive\end{tabular} & \begin{tabular}[c]{@{}c@{}}False\\ negative\end{tabular} \\
\midrule
\multirow{4}{*}{Baseline} 
& 0         & $10^{-5}$ & Yes  & -     \\
& $10^{-8}$ & $10^{-5}$ & Yes  & Yes   \\
& $10^{-5}$ & $10^{-2}$ & Yes  & Yes   \\
& $10^{-2}$ & $10^{-1}$ & -    & Yes   \\
\midrule
\multicolumn{3}{l}{\system tolerance estimation}  & - & -  \\
\bottomrule
\end{tabular}
\vspace{-0.5cm}
\end{table}

\subsection{Efficacy of Numerical Error Estimation}
\label{sec:eval:efficacy}

We address RQ4 by demonstrating \system's numerical error tolerance estimation accurately capture the characteristics of the numerical errors in distributed training, and that the numerical errors can indeed be distinguished from  actual bugs.
In \Cref{fig:fp-bug}, the numerical error (estimation) curves are obtained from the tolerance estimation in \Cref{sec:design:err-est}. The numerical error (distributed) curves are produced by the difference between the reference and a correctly implemented context parallel candidate.
In both BF16 and FP8 scenarios, the proximity between the estimation and the real numerical errors indicates our method can accurately reflect the behaviors of numerical errors.
For both data types, the bug-induced errors are approximately two orders of magnitude larger (100$\varepsilon$) than both the estimated and observed numerical ($\approx\varepsilon$).
This significant margin demonstrates that \system's dynamic error tolerance can effectively distinguish bug-induced errors from numerical errors.

\section{Related Work}
\label{sec:related}

\paragraph{ML system correctness and verification.}
Extensive research targets ML system correctness through testing frameworks~\cite{zhang_survey_2024, traincheck} and compilers~\cite{shen_comprehensive_2021,ma_survey_2023} using API fuzzing~\cite{cradle,audee,lemon,freefuzz,nablafuzz,zhang_duo_2021,zhang_predoo_2021} or graph generation~\cite{tzer,nnsmith,lin_deepdiffer_2023,wang_mlirsmith_2023,ma_fuzzing_2023} with
differential testing being a popular building block.
Parallel to testing, there is increasing popularity in techniques like verified graph rewrites or compilation~\cite{jia_taso_2019,arora_tensorright_2025,liu_verified_2024}, and equivalence analysis for computational graphs (e.g., Scalify~\cite{uva-egraph}, GraphGuard~\cite{graphguard}), which aim to prove semantic consistency in the compilation or parallelization process.
However, verification methods struggle to deal with bugs that exist on a lower level than their abstractions and the high maintenance cost of the bridge between implementation and specification. 
The rewrite-based verification method also needs to handle the cost of adding new rules and making sure the rules are correct.
Sharing a similar purpose to our work, TrainCheck\cite{traincheck} synthesizes variants from distributed execution traces and use them to find silent bugs in training.
Using a different approach, \system relies on a reference implementation to perform fine-grained differential testing on execution traces to find bugs.
It is also worth mentioning that Universal Checkpointing (UCP)~\cite{ucp} models distributed semantics for state consolidation rather than debugging, but its spirit is similar to our canonical tensor mapping procedure.

\paragraph{FP semantics.}
As low-bit FP data types like FP16 becomes more popular, there are emerging works characterizing the numerical effects of FP operations.  On the theory side, there are works
studying distribution of errors or bounds on the errors induced by FP operations~\cite{err-distr,inner-prod,sqrtn} from a probabilistic point of view, often with simplifying
assumptions on the actual FP hardware behaviors. More rigorous
modeling of the FP semantics~\cite{goubault_static_2001,solovyev_rigorous_2018,das_scalable_2020} is possible
yet expensive, making it challenging to scale to practical models.
Our work does not directly require precise modeling of FP differences, yet our technique and
theoretical derivations assume numerical stability of the single-device
and distributed FP implementation of neural network computations.

\paragraph{Sensitivity, smoothness, and robustness.}
Lipschitz properties characterize how the output of a neural network changes in response to input perturbations \cite{scaman_lipschitz_2019,qi_understanding_2023}. It is a desirable property to ensure training stability~\cite{qi_understanding_2023} and robustness against adversarial attacks~\cite{tsuzuku2018lipschitzmargintrainingscalablecertification}. In the context of transformers, it is well-known that the standard dot-product transformer is not Lipschitz continuous~\cite{kim_lipschitz_2021}, yet we can still derive bounds on Lipschitz constants of transformer layers
assuming compactly supported input~\cite{kim_lipschitz_2021,castin_how_2024} or local smoothness properties~\cite{havens_fine-grained_2024}. There are also works studying
Lipschitz continuous variants of transformers~\cite{kim_lipschitz_2021,qi_lipsformer_2023,havens_fine-grained_2024}.
Our work does not aim to perform a comprehensive study on Lipschitz constants
in neural networks, yet we rely on the ideal
and FP implementations of neural networks having certain smoothness properties, which
makes it possible to distinguish between expected FP differences and incorrect computations.

\section{Conclusion}
We present \system, the first system designed to systematically detect and localize \bugs in diverse practical training settings. 
\system collects intermediate tensors from distributed training in a fine-grained manner and compares them against those from a trusted single-device reference implementation. 
We conduct a principled study of how to distinguish FP round-off errors from bug-induced errors in distributed training, and propose a novel thresholding method grounded in theoretical analysis. 
\system detected 14 \bugs, including 3 new ones, in the widely used Megatron-LM framework.

\clearpage
\newpage
\bibliographystyle{plain}
\bibliography{reference}

\newpage
\onecolumn

\end{document}